\documentclass[aps,prb,preprint,superscriptaddress,longbibliography]{revtex4-1}

\usepackage{amsmath,amssymb}
\usepackage{graphicx}
\usepackage{epstopdf}
\usepackage{bm}
\usepackage{xcolor,ulem}

\newcommand{\CAO}     {CuAlO$_2$}
\newcommand{\al}	{$^{27}$Al}
\newcommand{\cu}	{$^{63}$Cu}
\newcommand{\slr} 	{$T_1^{-1}$}
\newcommand{\slrt} 	{$(T_1T)^{-1}$}

\newcommand{\kk} 	{$\mathcal{K}$}

\newcommand{\bc}[1]{\textbf{\sffamily #1}}

\begin{document}

\title[]{Observation of gapless collective charge fluctuations in an Anderson insulating state}

\author{Jong Mok Ok}
%\email[]{okjongmok@pusan.ac.kr}
\affiliation{Department of Physics, Pusan National University, Busan 46241, Korea}
\author{Beom Jun Park}
%\email[]{parkbeomjun2@gmail.com}
\affiliation{Department of Physics, Changwon National University, Changwon 51139, Korea}
\author{Junik Hwang}
%\email[]{wnsdlr0217@gmail.com}
%\homepage[]{Your web page}
\affiliation{Department of Physics, Changwon National University, Changwon 51139, Korea}
\author{Seonghoon Park}
%\email[]{digh10005@gmail.com}
%\homepage[]{Your web page}
\affiliation{Department of Physics, Changwon National University, Changwon 51139, Korea}
\author{Myeongjun Kang}
%\email[]{maudwnsm@pusan.ac.kr}
\affiliation{Department of Physics, 46241, Busan 46241, Korea}
\author{Jun Sung Kim}
%\email[]{js.kim@postech.ac.kr}
\affiliation{Department of Physics, Pohang University of Science and Technology, Pohang 790-784, Korea}
\author{Ki-Seok Kim}
\email[]{tkfkd@postech.ac.kr}
\affiliation{Department of Physics, Pohang University of Science and Technology, Pohang 790-784, Korea}

\author{Seung-Ho Baek}
\email[]{sbaek.fu@gmail.com}
%\homepage[]{Your web page}
\affiliation{Department of Physics, Changwon National University, Changwon 51139, Korea}
%\affiliation{Department of Materials Convergence and System Engineering, Changwon National University, Changwon 51139, Korea}

\date{\today}

%%%%%%%%%%%%%%%%%%%%%%%%%%%%%%%%%%%%%%%%%%%%%%%%%%%%%%%%%%%%%%%%%%%%

\begin{abstract}
\bf

Understanding the nature of collective charge dynamics in the Coulomb gap phase is essential for revealing the existence of many-body localization. However, the corresponding many-particle excitation spectra remain poorly understood.
Here, we present a comprehensive investigation of \al\ and \cu\ nuclear magnetic/quadrupole resonance (NMR/NQR), along with specific heat ($C_p$) measurements, in the $p$-type semiconductor \CAO. Our study unveils distinct changes in charge dynamics at two crossover temperature scales which separate three regimes associated with Anderson localization of charge carriers:  thermally activated transport ($T>150$ K) $\rightarrow$ Mott variable-range hopping (VRH) $\rightarrow$ Efros-Shklovskii (ES) VRH with Coulomb gap formation ($T<50$ K). In the ES VRH regime, we observe a striking divergence in the zero-field \cu\ spin-lattice relaxation rate, \slrt, 
which is strongly suppressed by an applied magnetic field, indicative of quantum critical charge fluctuations. This is further supported by a distinct magnetic field-dependence of $C_p/T$ deep within the Coulomb gap phase. Taken together, these results provide compelling evidence for the emergence of strong, gapless collective charge fluctuations within the Anderson insulating phase where single-particle excitations are gapped.

\end{abstract}

%\keywords{Nuclear magnetic resonance, charge-density-wave, square net}

\maketitle

%%%%%%%%%%%%%%%%%%%%%%%%%%%%%%%%%%%%%%%%%%%%%%%%%%%%%%%%%%%%%%%%%%%%
 
Anderson localization of electrons \cite{anderson58a} has been a topic of extensive research for over half a century, particularly in semiconductors hosting impurities or defects, revealing a variety of intriguing transport phenomena \cite{evers08,abrahams10,dobrosavljevic12}. Assuming the existence of a mobility edge $\epsilon_\text{M}$ located below the Fermi energy $\epsilon_\text{F}$, three distinct transport regimes emerge, characterized by two crossover temperature scales. At elevated temperatures ($T> \epsilon_\text{F} - \epsilon_\text{M}$), electrical conductivity $\sigma(T)$ follows an activated behavior, $\sigma(T) \propto \exp[-(\epsilon_\text{F} - \epsilon_\text{M})/T]$. Upon cooling below a crossover temperature $T_\text{M}$,  charge transport occurs via hopping between localized states, as described by Mott variable-range hopping (VRH), with $\sigma(T)\propto \exp[-(T_M/T)^{\frac{1}{d+1}}]$, where $d$ denotes the spatial dimensionality of the system \cite{mott69, apsley74}. 
At even lower temperatures, below a second crossover temperature $T_\text{ES} < T_\text{M}$, Coulomb interactions between localized states lead to the formation of a soft gap in the electronic density of states (DOS) near the Fermi level, the so-called Coulomb gap.  The Coulomb gap modifies the VRH conduction law, changing the exponent from $1/(d+1)$ to 1/2 for all spatial dimensions, thereby defining the Efros-Shklovskii (ES) VRH regime \cite{efros75,rosenbaum91}. %The crossover from Mott to Efros-Shklovskii VRH has been observed in various materials \cite{rosenbaum91,aharony92,moon18,kumar24}. 

While single-particle dynamics in this regime are well understood \cite{efros75}, many-body dynamics within the Coulomb gap phase remains poorly characterized. Most theoretical efforts to date have focused on the phenomenon of \textit{many-body localization}, which is predicted to occur in the weak Coulomb interaction limit \cite{gornyi05,basko06,nandkishore15}. In contrast, an alternative theoretical framework based on self-thermalization in disordered,  interacting electron systems (in the absence of  a phonon bath) predicts the emergence of gapless collective charge fluctuations, even when single-particle excitations are localized \cite{mueller07,andreanov12}. These gapless modes could potentially lead to \textit{many-body delocalization} \cite{mueller07}, a phenomenon that remains experimentally unconfirmed.
In this work, we employ nuclear magnetic/quadrupole resonance (NMR/NQR) and specific heat measurements to investigate the $p$-type semiconductor \CAO, extending the previous NMR/NQR studies \cite{ogloblichev18,ogloblichev21} conducted in polycrystalline forms, and  uncover the emergence of gapless collective charge fluctuations within the Coulomb gap phase. Intriguingly, these fluctuations are strongly suppressed under a moderate external field of 5.716 T, accompanied by the development of spatial inhomogeneity that is absent in zero field. This observation provides new insight into the nature of many-body delocalization in the Coulomb gap regime, highlighting the interplay between magnetic field, charge dynamics, and localization phenomena.

The inset of Fig.\,1a shows the crystal structure of \CAO\ (space group: $R\bar{3}m$), characterized by alternating layers of close-packed two-dimensional Cu planes and edge-sharing AlO$_6$ octahedra, stacked along the $c$-axis. These layers are interconnected via O-Cu-O dumbbell-like bridges (highlighted in yellow) \cite{shannon71}. \CAO\ is a well-known $p$-type semiconductor with a band gap of 3.5 eV and exhibits optical transparency in thin film form \cite{kawazoe97, yanagi00}. The $p$-type conductivity is generally attributed to Cu vacancies, which can give rise to Cu$^{2+}$ centers \cite{katayamayoshida03,raebiger07,tate09,scanlon10a}, with the associated acceptor level estimated to lie about 0.7 eV above the valence band edge \cite{tate09,scanlon10a}.  

Figure 1a presents the quadrupole-perturbed \al\ NMR spectra acquired at 120 K for external magnetic fields applied both parallel ($H \parallel c$) and perpendicular ($H \perp c$) to the $c$-axis.
The observed four satellite transitions, corresponding to the $m \leftrightarrow m-1$ transitions ($m=-I,-I+1,\ldots, +I$) are well-described by the expression \cite{bennet}: $\frac{1}{2}\nu_Q(3\cos^2\theta -1)\left(m-\frac{1}{2}\right)$
where $\nu_Q$ is the nuclear quadrupole frequency and $\theta$ is the angle between the principal axis $z$ of the electric field gradient (EFG) and the applied magnetic field $\mathbf{H}$, confirming axial symmetry at the Al site. The central transitions ($\frac{1}{2}\leftrightarrow -\frac{1}{2}$) for both field directions are shown separately in Fig.\,1b.  
Remarkably, all \al\ lines exhibit exceptional sharpness, with full widths at half maximum (FWHM) less than 10 kHz (see Fig.\,1e), providing evidence for the high crystalline quality of the sample.
For \cu\ NMR, only the central lines were measured (Fig.\,1c) due to the large $\nu_Q$,  which was independently determined from \cu\ NQR measurements (inset of Fig.\,4a).

The temperature dependence of the Knight shift \kk\ and FWHM for both \al\ and \cu\ spectra is presented in Figs.\,1d and 1e, respectively, for the two field orientations ($H\parallel c$ and $H\perp c$).   
The nearly constant values of both \kk\ and FWHM across the full temperature range indicate that potential local magnetic moments associated with Cu$^{2+}$ impurities, estimated to occupy about 1.4\% of Cu sites \cite{kim22}, have negligible influence on the NMR spectra.
Remarkably, all dataset exhibit subtle yet discernible variations below $T_0=50$ K. In particular, the simultaneous increase in FWHM for both \al\ and \cu\ spectra suggests the development of spatial inhomogeneity across the sample below $T_0$. Given the absence of local moment effects, we attribute this anomaly to the onset of charge interactions that induce local variations in the EFG, manifesting as line broadening. 
As discussed further below, this spatial inhomogeneity is likely due to the coexistence of localized and delocalized charge carriers in the presence of an external magnetic field.

Figures 2a and 2b show the temperature dependence of the spin-lattice relaxation rate \slr\ for \al\ and \cu\ nuclei, respectively, while Figs.\,2c and 2d display the temperature dependence of \slrt, which provides additional insight into low-energy spin and charge dynamics. 
For both nuclei, the \slr\ data exhibit prominent peaks near 200 K, well captured by the Bloembergen, Purcell, and Pound (BPP) model \cite{bloembergen48},
\begin{equation}
	T_1^{-1} \propto \frac{\tau_c}{1+\tau_c^2\omega_L^2},
	\label{bpp}
\end{equation}
where $\omega_L$ is the nuclear Larmor frequency, and $\tau_c=\tau_c^0\exp(E_a/T)$ is the correlation time of local fluctuations, with $E_a$ representing the activation energy and $\tau_c^0$ the high-temperature limit of the correlation time. A small constant Korringa-like term, $(T_1T)^{-1}=0.012$ (sK)$^{-1}$, is added to account for the weak non-BPP contribution at high temperatures. 
The BPP model describes a peak in \slr\ that arises when a continuous slowing down of $\tau_c$ for certain local fluctuations crosses the inverse of $\omega_L$, i.e., $\tau_c\omega_L=1$. 
In the case of \CAO, a $p$-type semiconductor, $\tau_c$ can be interpreted as the hopping time of thermally activated holes. Fitting the data to this model with a fixed $\tau_c^0=1\times 10^{-11}$ s yields activation energies of $E_a =0.09$ eV and 0.084 eV for $H\parallel c$ and $H\perp c$, respectively.
Interestingly, these values are in good agreement with the estimate of approximately 0.1 eV reported in the previous NMR study\cite{ogloblichev21}, although we note that a full BPP fit was not successfully achieved in that study.
These excellent BPP fits strongly suggest that, at high temperatures ($T>150$ K), charge transport in \CAO\ is dominated by thermally activated hopping of small polarons\cite{benko84,ingram01}, rather than by band conduction \cite{tate09}. 
Furthermore, as shown in Figs.\,2c and 2d, the \slrt\ data for both \al\ and \cu\ nuclei collapse onto a single curve after rescaling the \al\ data, indicating that the relaxation mechanisms are governed by a common fluctuation spectrum. This behavior implies that hole hopping occurs not only the Cu layers but also between Cu and AlO$_6$ layers, reflecting a three-dimensional nature of charge dynamics \cite{buljan99,lee01a}. 
Below 150 K, the temperature dependence of \slr\ begins to deviate noticeably from the BPP behavior, signaling the onset of an additional relaxation mechanism. %\shb{\sout{associated with a new electronic phase.} } 
Interestingly, this deviation coincides with a crossover in the electrical conductivity from thermally activated behavior to Mott VRH \cite{kawazoe97,yanagi00,lee01a}. 
A plausible mechanism for this enhanced nuclear relaxation in the Mott VRH regime involves electron tunneling between localized states. Each tunneling event can abruptly modify the local charge distribution, leading to quadrupole relaxation of nearby nuclei. In the strong collision limit, such events can be highly efficient in inducing nuclear transitions \cite{alexander65,baek05}, thereby enhancing \slr. 

The most strking feature in Figs.\,2c and 2d is a sharp upturn in \slrt\ below $T_0=50$ K, a temperature that also marks the onset of changes in both \kk\ and spectral linewidth (Figs.\,1d and 1e). Since local moments are not evident from the NMR data, the observed upturn in \slrt\ is attributed to low-energy charge fluctuations. To unambiguously identify the origin of the relaxation mechanism, we measured \slrt\ for $^{65}$Cu with $H \perp c$. If spin fluctuations dominate the relaxation process, the ratio of \slrt\ between the two Cu isotopes should scale as $(^{63}\gamma/^{65}\gamma)^2$, where $\gamma$ is the nuclear gyromagnetic ratio. In contrast, if charge fluctuations dominate, the ratio should follow $(^{63}Q/^{65}Q)^2$, where $Q$ denotes the nuclear quadrupole moment. As shown in Fig.\,3, the observed ratio of \slrt\ clearly matches the latter scenario, providing direct evidence that charge fluctuations are the dominant relaxation mechanism in the VRH regime below  $\sim$150 K.

To further explore the role of an external magnetic field on charge fluctuations, we performed \cu\ \slr\ measurements using NQR, which probes \slr\ in zero magnetic field.  Figure 4a compares the temperature dependence of \cu\ \slrt\ obtained from NQR with that measured via NMR at 5.716 T for $H\perp c$. Remarkably, the upturn in \slrt\  below $T_0$ is nearly an order of magnitude larger in the zero-field NQR data than in \cu\ NMR results.
Additionally, the \cu\ NQR spectrum remains exceptionally sharp over the entire temperature range (inset of Fig.\,4a), in stark contrast to the clear line broadening seen in all NMR spectra below $T_0$ (Fig.\,1e).
This discrepancy implies that the application of a magnetic field not only suppresses charge fluctuations but also induces a spatial inhomogeneity below $T_0$. The underlying origin of this unusual magnetic field effect will be discussed below.

Having established the development of strongly field-dependent charge dynamics below $T_0$, it is desirable to explore thermodynamic properties to elucidate their origin. To this end, we investigated the specific heat $C_p$ as a function of temperature and magnetic field, as shown in Fig.\,4b. 
In zero magnetic field, we find that $C_p/T$ follows a power-law behavior $C_p/T\propto T^{3.1}$ below approximately $T_0$, decreasing significantly faster than the Debye model at low temperatures $C/T\propto T^2$ (blue dashed line in Fig.\,4b). This fast suppression of $C_p/T$ suggests a reduction in the electronic DOS near the Fermi level, consistent with the formation of a pseudogap.
Interestingly, $C_p/T$ deviates from the power-law behavior around 12 K and exhibits an upturn below 4 K. Upon applying a magnetic field, the pseudogap-like suppression of $C_p/T$ persists up to 12 T. However,  the low-temperature enhancement of $C_p/T$ shows a nontrivial field dependence: it initially increases with field strength and is subsequently suppressed at higher fields.

At first glance, the low-temperature upturn in $C_p/T$ and its field dependence resemble a Schottky anomaly, possibly arising from magnetic contributions of Cu$^{2+}$ impurity ions. However, the zero-field $C_p/T$ data do not follow the high-temperature tail expected for a Schottky anomaly, i.e., $C_p/T\propto T^{-3}$, as indicated by the yellow dotted curve in Fig.\,4b. Furthermore, the field dependence of the enhancement---initially increasing and then decreasing---is clearly non-monotonic, which contrasts with the typical behavior of a Schottky anomaly, where monotonic broadening and shift of the anomaly with increasing magnetic field is expected \cite{rajan82}.
Instead, we propose that the observed low-temperature anomaly in $C_p/T$, together with the diverging \slrt\ detected via \cu\ NQR, reflects quantum critical dynamics. Specifically, these features suggest proximity to a quantum phase transition associated with low-energy collective excitations within the Coulomb gap phase at $T=0$. 
Further support for this proposal comes from the zero-field behavior of the NQR relaxation rate. As shown in Fig.\,4c, $T_1T$, which is proportional to the inverse of the $q$-summed imaginary part of the dynamical susceptibility at $\omega_L$, $1/\sum_q \chi''(q,\omega_L)$, approaches zero linearly as $T\rightarrow 0$. This behavior strongly implies that the divergence of \slrt\ observed in \cu\ NQR arises from quantum critical charge fluctuations, as identified in Fig.\,3 and discussed in the context of quantum critical phenomena \cite{barzykin09a}.
%

%%%%%
We now turn to the physical implications of our NMR/NQR and specific heat results in \CAO.
Within the framework of single-particle Anderson localization, our study reveals two notable crossover phenomena. Firstly, we identify a crossover from thermally activated transport to Mott VRH near 150 K, marking a transition into the regime of weak localization, or precursor of Anderson localization \cite{bergmann84}. This crossover is clearly manifested in the temperature dependence of the nuclear spin-lattice relaxation rate, which deviates from BPP behavior below this temperature.
Secondly, a more significant crossover is observed at $T_0\approx 50$ K, where our data indicate a transition from the Mott VRH regime to the ES VRH regime, corresponding to strong localization.  This is evidenced by the formation of a pseudogap in the electronic DOS, as inferred from the low-temperature power-law behavior $C_p/T\propto T^{3.1}$ (Fig.\,4b). Interpreting this pseudogap as a Coulomb gap, we infer that the DOS within the localized regime follows the form $\rho(\epsilon) \propto \epsilon^{2\alpha}$ with $\alpha\sim 1.5$. 
This exponent exceeds the mean-field prediction of $\rho(\epsilon) \propto \epsilon^2$ (i.e., $\alpha=1$) \cite{efros75}. The deviation suggests that strong charge fluctuations lead to a renormalization of the mean-field DOS.% (see the Supplementary Material for a more detailed analysis).

However, the divergent collective charge fluctuations detected via \cu\ NQR within the Coulomb gap phase (Fig.\,4a) are difficult to reconcile with the Anderson localization picture, which assumes the localization of single-particle excitations. This strongly suggests that electron-electron correlation effects become increasingly significant at low temperatures, necessitating a theoretical framework that extends beyond single-particle localization physics. 
Indeed, near the proposed zero-temperature Anderson transition, the localization length $\xi$ is expected to grow substantially \cite{efetov92}, enhancing long-range Coulomb interactions among localized electrons. These  interactions may enable many-body delocalization processes, even when the Anderson insulating state remains intact \cite{mueller07,andreanov12}.

Compelling support for this scenario is provided by the drastic suppression of the divergent enhancement in \cu\ \slrt\ at 5.716 T (Fig.\,4a). A magnetic field breaks time-reversal symmetry and can disrupt interference effects crucial for single-particle localization \cite{altshuler81}, potentially giving rise to a spatially inhomogeneous coexistence of localized and delocalized states. This field-induced inhomogeneity is directly evidenced by the line broadening of the NMR spectra below $T_0$ in an applied field, in contrast to the unchanged linewidth observed in zero-field \cu\ NQR spectra (see the inset of Fig.\,4a). 
These emergent delocalized states, induced by the magnetic field, can screen long-range Coulomb interactions and thereby suppress the electron-assisted many-body excitations responsible for the divergent \slrt\ enhancement.

Based on the self-consistent renormalization method in the one-loop level (TAP approach in ref.~\cite{mueller07} and replica symmetry breaking in ref.~\cite{andreanov12}), the calculation of the density-density correlation function in the quantum Coulomb glass model (or the spin-spin correlation function in the Ising quantum spin glass model) reveals the existence of gapless charge dynamics. However, the resultant spectral function exhibits a linear dependence on frequency in the low-frequency limit, which is inconsistent with the observed divergent collective charge fluctuations in the Coulomb gap phase. Certainly, further theoretical investigations, which are likely non-perturbative in nature, are warranted to fully comprehend the divergent charge dynamics developed in the Anderson localized state.

\subsection*{Methods}

High quality single crystals of \CAO\ were prepared by a reactive crucible melting method detailed in ref.\,\cite{kim22}. The sample was mounted on a goniometer to ensure precise alignment with the external field. 
\al\ (nuclear spin $I=5/2$) and $^{63,65}$Cu ($I=3/2$) NMR measurements were carried out at an external field of $H=5.716$ T. For \cu, we also measured NQR in zero field. The NMR/NQR spectra were acquired using a standard spin-echo technique with a typical $\pi/2$ pulse length 3-4 $\mu$s. The spin-lattice relaxation rate \slr\ was determined using a saturation method.  The recovery curves of the nuclear magnetization $M(t)$ were fitted to the appropriate fitting functions. For $^{63,65}$Cu NMR, 
$$1-M(t)/M(\infty) = A \left[0.1\exp(-t/T_1) +0.9\exp(-6t/T_1)\right],$$ 
for $^{27}$Al NMR,
  $$1-M(t)/M(\infty) = A \left[\frac{1}{35}\exp(-t/T_1) +\frac{8}{45}\exp(-6t/T_1)+\frac{50}{63}\exp(-15t/T_1)\right],$$  and for $^{63}$Cu NQR, $$1-M(t)/M(\infty) = A \exp(-3t/T_1),$$ where $A$ is a fitting parameter that is ideally one.

Specific heat measurements were performed using the Quantum Design Physical Properties Measurement System (PPMS).

\subsection*{Data Availability}

The data that support the findings of this study are available from the corresponding authors (K-S.K. and S-H.B.).

\subsection*{References}
%\bibliographystyle{naturemag}
%\bibliography{mybib}

%

\let\origdescription\description
\renewenvironment{description}{
  \setlength{\leftmargini}{0em}
  \origdescription
  \setlength{\itemindent}{0em}
  \setlength{\labelsep}{\textwidth}
}
{\endlist}

\begin{description}
\item[Acknowledgments] This research received funding from the `Mid-career Faculty Research Support Grant' at Changwon National University in 2025.
\item[Funding] This work was supported by the National Research Foundation of Korea (NRF) grants funded by the Korea Government (MSIT) (Grant Nos.~NRF-2020R1A2C1003817, NRF-2021R1A2C1006453, NRF-2021R1A4A3029839, NRF-2022R1A2C3009731,  RS-2023-00210295, and RS-2023-00221154). \
\item[Author Contributions] JMO, KSK, and SHB have proposed and initiated the project.  MJK and JMO have grown single crystals, and KSK carried out theoretical calculations. BJP, JH, SP, and SHB performed NMR measurements and analyzed data. JMO, KSK, and SHB participated in writing of the manuscript. All authors
		discussed the results and commented on the manuscript.
%\item[Competing financial interests] The authors declare no
%	competing financial interests.
\item[Additional information] Correspondence and requests for materials should be
	addressed to S.-H. Baek~(email: sbaek.fu@gmail.com).

\end{description}

\pagebreak

\begin{figure*}
\centering
\includegraphics[width=\linewidth]{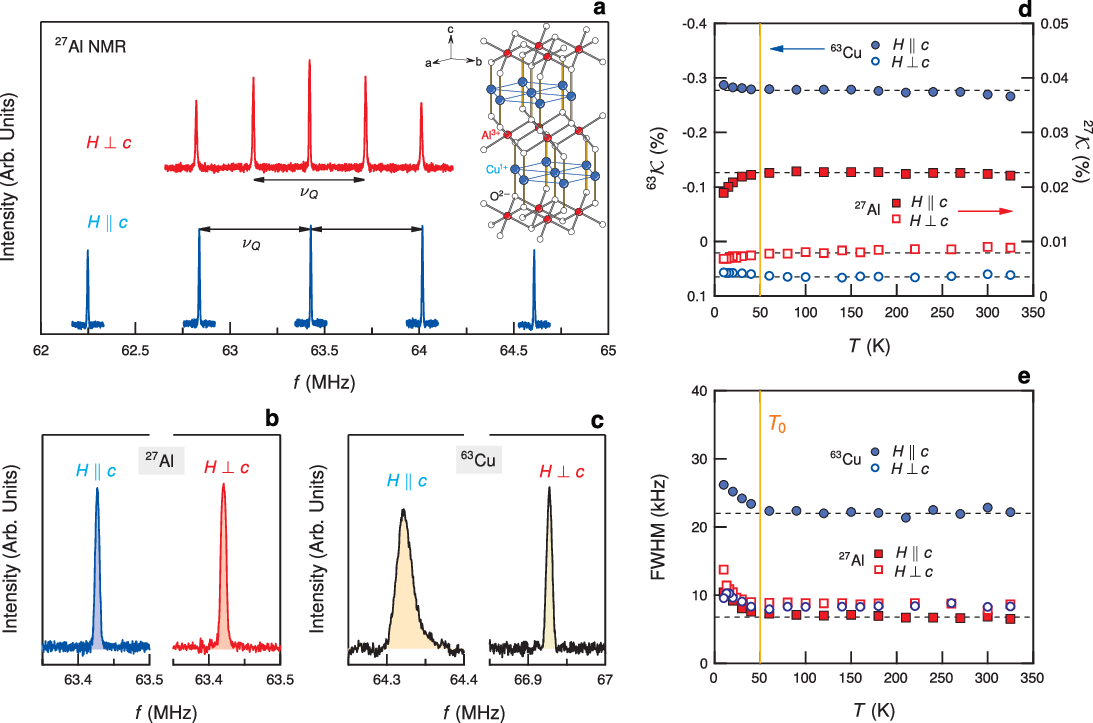}
\caption{ \bc{a}, \al\ ($I=5/2$) NMR spectrum measured at 120 K under an external field of 5.716 T for two field orientations. The symmetric satellite pairs reflect axial symmetry about the $c$ direction at the Al site, yielding a well-defined quadrupole frequency $\nu_Q$ for \al. The inset depicts the crystal structure of CuAlO$_2$.
  \bc{b}-\bc{c}, Central transition lines ($\frac{1}{2} \leftrightarrow -\frac{1}{2}$) of \al\ (\bc{b}) and \cu\ (\bc{c}). \bc{d}-\bc{e}, Temperature dependence of the Knight shift \kk\ (\bc{d}) and FWHM (\bc{e}) for the \al\ and \cu\ central transitions. Both \kk\ and FWHM remain nearly constant down to $T_0=50$ K for both nuclei.
  }
\label{spec}
\end{figure*}

\begin{figure}
\centering
\includegraphics[width=0.65\linewidth]{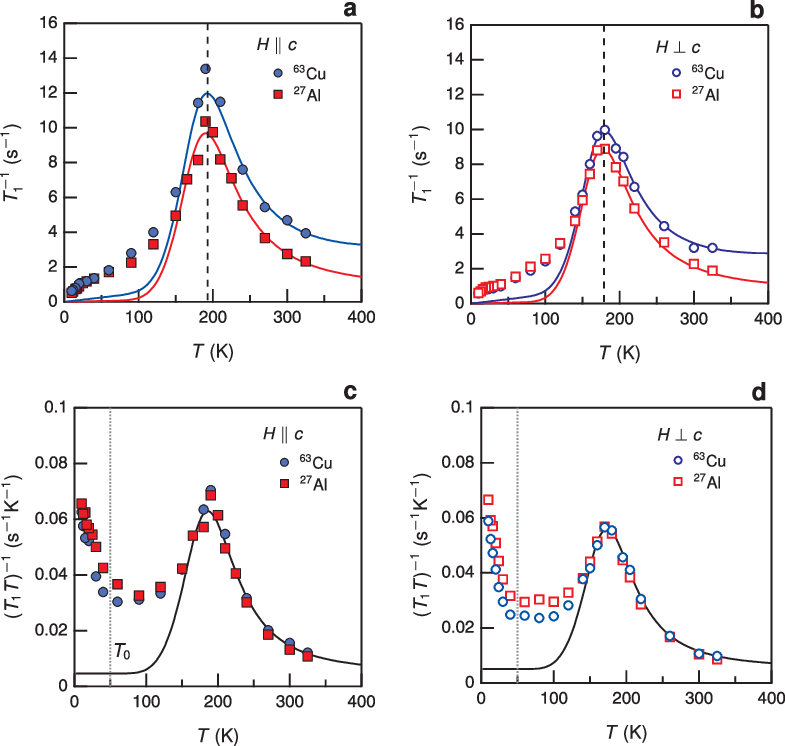}
\caption{\bc{a}-\bc{b}, Temperature dependence of \slr\ for \al\ and \cu\ with the external magnetic field applied parallel (\bc{a}) and perpendicular (\bc{b}) to the $c$-axis. 
\bc{c}-\bc{d}, Temperature dependence of \slrt\ for $H\parallel c$ (\bc{c}) and $H\perp c$ (\bc{d}). The \slrt\ data for \al\ have been rescaled to match those of \cu.
The peak near 200 K is well described by the BPP model for both nuclei (solid lines), yielding an activation energy of approximately 0.09 eV. Below $\sim150$ K, the \slr\ data deviate from BPP behavior. A pronounced upturn in \slrt\ is observed below $T_0=50$ K for both nuclei and both field orientations.
}
\label{t1t}
\end{figure}

\begin{figure}
\centering
\includegraphics[width=0.5\linewidth]{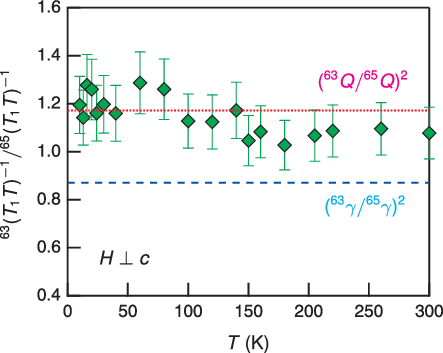}
\caption{The \slrt\ ratio between the \cu\ and $^{65}$Cu isotopes matches the square of the ratio of their nuclear quadrupole moments $(^{63}Q/^{65}Q)^2$ at low temperatures, indicating that charge fluctuations dominate the nuclear spin-lattice relaxation process.
}
\label{ratio}
\end{figure}

\begin{figure*}
\centering
\includegraphics[width=\linewidth]{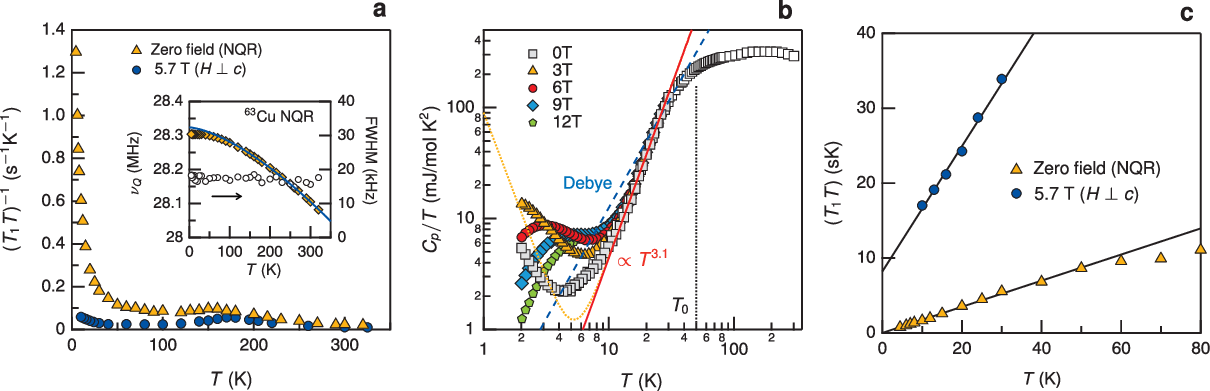}
\caption{ \bc{a}, Temperature dependence of \slrt\ measured via \cu\ NQR in zero field. A significantly stronger enhancement of \slrt\ is observed below $T_0$, by nearly an order of magnitude, compared to the \cu\ NMR results obtained at 5.716 T. The inset shows $\nu_Q$ (left axis) and FWHM (right axis) of the \cu\ NQR spectrum as  functions of temperature. The solid line represents a phenomenological $T^{3/2}$ dependence, consistent with a phonon contribution. %The FWHM remains narrow without any discernible temperature dependence. 
\bc{b}, Specific heat divided by temperature, $C_p/T$, plotted as a function of temperature and magnetic field. A power law behavior, $C_p/T\propto T^{3.1}$, is observed below $T_0$. For comparison, the Debye model prediction $C_p/T\propto T^2$ is drawn as a blue dashed line, and the yellow dotted line represents the expected Schottky anomaly in zero field. %The data deviate from the pseudogap behavior below $\sim 15$ K, being strongly dependent on $H$. 
\bc{c}, Temperature dependence of $T_1T$ from both \cu\ NQR and NMR measurements. In zero field, $T_1T$ approaches zero as $T\rightarrow 0$, indicating the emergence of quantum critical charge fluctuations.
}
\label{NQR}
\end{figure*}

%%%%%%%%%%%%%%%%%%%%%%%%%%%%%%%%%%%%%%%%%%%%%%%%%%%%%%%%%%%%%%%%%%%%

\end{document}